# Enzymatic AND Logic Gate with Sigmoid Response Induced by Photochemically Controlled Oxidation of the Output

Vladimir Privman,[a] Brian E. Fratto,[b] Oleksandr Zavalov,[a] Jan Halámek,[b] and Evgeny Katz[b]

[a]Department of Physics, and [b]Department of Chemistry and Biomolecular Science, Clarkson University, Potsdam, NY 13676

**ABSTRACT:** We report a study of a system which involves an enzymatic cascade realizing an **AND** logic gate, with an added photochemical processing of the output allowing to make the gate's response sigmoid in both inputs. New functional forms are developed for quantifying the kinetics of such systems, specifically designed to model their response in terms of signal and information processing. These theoretical expressions are tested for the studied system, which also allows us to consider aspects of biochemical information processing such as noise transmission properties and control of timing of the chemical and physical steps.

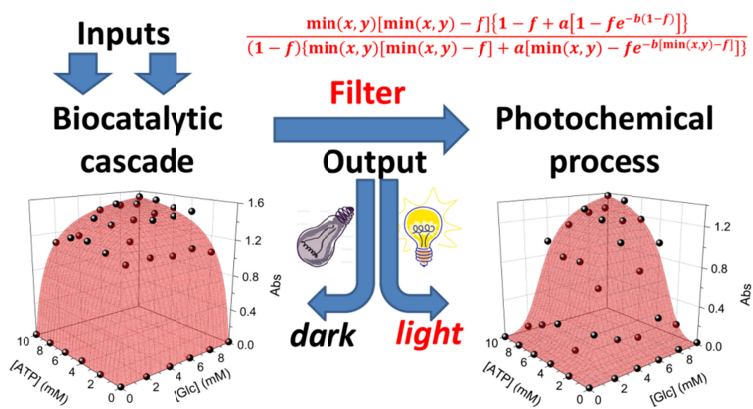

**Keywords:** Unconventional computing, Biomolecular computing, Enzyme computing, Noise, NAD$^+$ recycling.

## 1. Introduction

Molecular[1] and biomolecular[2] information processing systems have been extensively investigated in recent years in the framework of unconventional computing[3,4] and biosensing.[5-7] Chemical[8-12] and biochemical[13-17] systems mimicking binary logic gates and their networks have been studied for novel computational and signal processing designs. Biomolecular systems inspired by natural processes include a variety of species of different complexity including systems based on proteins/enzymes,[17,18] DNA,[16,19] RNA[20] and whole cells.[21-23] Scaling up the complexity of chemical and particularly biochemical logic systems resulted in the formulation of systems performing several operations and sophisticated functions. For example, enzyme-based systems were assembled in multi-step biocatalytic cascades performing several logic operations[24,25] and arithmetic functions.[26] Realization of the negation function in biocatalytic reactions allowed construction of the universal NAND / NOR gates.[27] Computation based on chemical kinetics was applied for mimicking biological neuron networks.[28-32]

While computation based on biomolecules[33] is a futuristic concept, short-term applications of this research include binary-decision alert-type sensor systems.[5-7] Biomolecular computing systems can operate in biological and biotechnological environments[34] and they can therefore be utilized for multiplexing and processing multiple biochemical signals in the binary format, **0** and **1**, relying on (bio)chemical processes rather than electronics.[5-7] Biomolecular logic systems have been explored for biomedical/diagnostic applications aiming at analysis of biomarkers characteristic of various pathophysiological conditions of interest in diagnostics of diseases[35-37] or injuries.[38-43] They have also been considered for analysis of explosives and toxic materials,[44] as well as radiation effects.[45]

Achieving versatile and multifunctional information processing with biomolecules in the systems offering new functionalities supplementary to electronics, has been the primary challenge in the field.[2,17,46-49] For multi-step, multi-signal information processing of increased levels of complexity, binary gates have to be connected in networks.[24,48-53] This requires development of paradigms for scalability and avoidance of noise buildup,[48,49] as well as "clocking" (temporal control) and ultimately spatial separation of the various steps[52,53] of multistage biochemical processes. Furthermore, biochemical processes have to be interfaced[7,17,54] with physical signals for input, output, and control. Ideas borrowed from the binary/digital information processing approaches in electronics have proven useful. Specifically, recently several approaches[17,47,55-63] have been proposed and realized, utilizing simple added chemical steps to convert the response of enzymatic processes which mimic binary gates to sigmoid. This allows suppression of noise levels, by what amounts to signal filtering,[47,55-65] during transmission/processing from the inputs of the gate to its output.

In this work we consider a model system which combines enzyme-catalyzed biochemical steps with an optical physical-signal activated chemical process yielding filtering which makes the system's response sigmoid in both inputs. This system realizes an **AND** logic gate, and offers interesting properties for a study of aspects of biochemical information processing. Specifically, it involves "clocking" by an approximate time-scale separation (fast vs. slow enzymatic processes) and by external (optical) control. It also illustrates aspects of possible response patterns of two-input, one-output **AND** gates, including filtering to achieve noise suppression.

The system is used as a test case to develop and apply kinetic expressions specifically devised for phenomenological data fitting suitable for evaluation of the quality as a logic gate, for processing information and for incorporation in networks. There has been a discussion[24,48,66-68] in the literature of the suitability of phenomenological kinetic expressions, notably the Hill equation,[66-68] for fitting data for enzyme-catalyzed and other processes with sigmoid response. Our present work introduces a few-parameter description which is particularly suitable for the afore-cited[47,55-63] recent approaches to "biochemical filtering" by adding chemical steps, rather than the traditional mechanisms for sigmoid response resulting from cooperativity. We note that the latter mechanisms, involving, for instance, enzyme allostericity, have also been considered[64,65,69,70] in studies aimed at applications involving biomolecular information and signal processing.

## 2. Experimental Section

*Chemicals and Materials.* The following enzymes were obtained from Sigma-Aldrich and used without further purification: hexokinase (HK) from *Saccharomyces cerevisiae* (E.C.2.7.1.1), glucose-6-phosphate dehydrogenase (G6PDH) from *Leuconostoc mesenteroides* (E.C. 1.1.1.49). Other chemicals from Sigma-Aldrich used as supplied included: α-D-glucose (Glc), adenosine 5′-triphosphate disodium salt hydrate (ATP), β-nicotinamide adenine dinucleotide sodium salt (NAD$^+$) and *tris*(hydroxymethyl)aminomethane (Tris-buffer). Thionine acetate was obtained from Alfa Aesar. Ultrapure water (18.2 MΩ cm) from NANOpure Diamond (Barnstead) source was used in all of the experiments.

*Instrumentation and Measurements.* A Shimadzu UV-2401PC/2501PC UV-Vis spectrophotometer (Shimadzu, Tokyo, Japan) with 1 mL poly(methyl methacrylate) (PMMA) cuvettes was used for all measurements. Measurements were taken in a thermostated cuvette (20 ± 0.5 °C) and all reagents were brought to this temperature prior to measurements. The halogen bulb lamp (500 W T3 halogen bulb) was purchased from ACE Hardware Corp. During the photochemical activation, the reaction solution was illuminated with non-filtered polychromatic light from the lamp with the intensity controlled by applying variable voltage to the lamp. The light intensity was measured with Light Meter LX802 (MN Measurement Instruments).

*Composition and Operation of the System.* The biocatalytic cascade performing the **AND** gate operation was catalyzed by the two enzymes, HK (1 unit mL$^{-1}$) and G6PDH (1 unit mL$^{-1}$) operating in the presence of NAD$^+$, Glc and ATP. Note that oxygen was present in the solution under equilibrium with air. The system was realized in 0.1 M Tris-buffer, pH 7.3, with Mg(CH$_3$COO)$_2$ (6.7 mM), used also as the reference background solution for all optical measurements. NAD$^+$ was always added at the initial concentration of 0.25 mM, while Glc and ATP were selected as inputs activating the **AND** logic gate. The logic-**0** and **1** values for the two inputs were selected as 0 and 10 mM, respectively. For mapping the **AND** gate response function the input concentrations were varied between 0 and 10 mM in 2 mM increments, resulting in the data matrix of 6×6 points with all combinations of the input concentrations. After the biocatalytic reaction was performed for 600 s, resulting in the production of NADH limited by the input concentrations, thionine was added (200 or 250 µM) and the solution was illuminated for 300 s with the non-filtered light from the halogen lamp with intensity of 9000 or 12000 lux. The NADH concentration was measured optically at $\lambda = 340$ nm (extinction coefficient 6.22 $M^{-1}cm^{-1}$)[71] before and after this photochemical activation step of the experiment representing the output signal without and with "filtering" process, respectively.

## 3. The System and Its Functioning as an AND Logic Gate

The biocatalytic processes occurring in the presence of all reacting species are sketched in Scheme 1. Hexokinase (HK) catalyzes phosphorylation of one of the input chemicals, glucose (Glc) in the presence of the other input, ATP, yielding glucose-6-phosphate (Glc6P).[72] In order to convert the resulting product to the desired measurable signal, glucose-6-phosphate dehydrogenase (G6PDH) catalyzes the oxidation of Glc6P to yield 6-phosphogluconic acid (6-PGluc) with the concomitant reduction of $NAD^+$ to NADH.[72,73] The biocatalytically produced NADH represents the final output signal generated by the system, read out optically at $\lambda = 340$ nm. It should be noted that both input species (Glc and ATP) react stoichiometrically in the 1:1 ratio and therefore the amount of the final product (NADH) is limited by the minimum amount of Glc or ATP initially provided. In order to realize the "filter" operation, NADH was "recycled" in the controlled way back to $NAD^+$.[55,57] Such recycling, which is an important process in many biotechnological systems, e.g., in bioreactors,[74,75] biofuel cells,[76,77] and biosensors,[78,79] can be performed by biocatalytic,[80,81] electrocatalytic[82,83] or photocatalytic oxidation.[84-87] Here we used photochemical oxidation by photoexcited thionine dye in the presence of oxygen acting as an electron acceptor, Scheme 1.

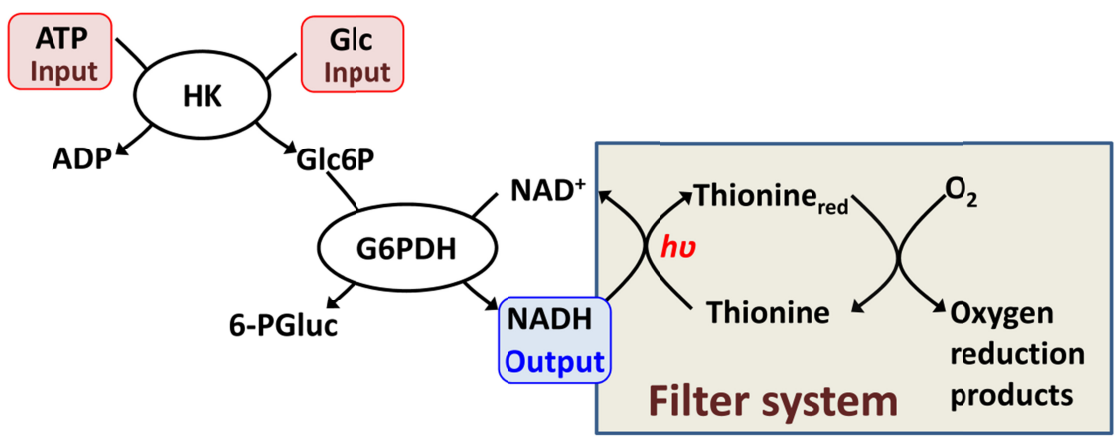

**Scheme 1.** The biocatalytic cascade for processing biomolecular input signals mimicking **AND** logic gate, followed by a photochemically activated "filter" process for converting a convex-shape response function to sigmoid. The abbreviations stand for: ATP = adenosine 5′-triphosphate, ADP = adenosine 5′-diphosphate, Glc = α-D-glucose, Glc6P = glucose-6-phosphate, 6-PGluc = 6-phosphogluconic acid, $NAD^+$ = β-nicotinamide adenine dinucleotide, NADH = β-nicotinamide adenine dinucleotide reduced, HK = hexokinase, G6PDH = glucose-6-phosphate dehydrogenase. The photochemical action of thionine is shown in a simplified schematic version.

The photochemical process resulting in the oxidation of NADH and reduction of $O_2$ is multistep, including photoexcitation of thionine to its higher singlet state, then conversion to the triplet state followed by the triplet state redox reaction to yield intermediate radical species finally regenerated to the initial ground state of thionine with concomitant oxidation of NADH and reduction of oxygen.[88] In a very simplified description, the process could be represented as a photocatalytic reaction mediated by thionine which is activated by light. Similar processes, including their various mechanisms and intermediate species were described in the literature;[89,90] their detailed description is outside the scope of the present study. Because of the complexity of the photochemical process, including generation of various intermediate species and various products of the oxygen reduction, the process is included in an oversimplified way in Scheme 1, which shows the photochemical pathway as the reaction of the photoexcited thionine with NADH,[88] resulting in the reduced state of thionine. It should be noted, however, that another pathway,[90] where the photoexcited thionine reacts with oxygen yielding the oxidized state of thionine, is also possible. If this oxidative pathway is realized, the oxidized state of thionine will react with NADH (instead of the photoexcited state shown in the scheme) resulting in the formation of $NAD^+$. However, the photochemical cycle will result in the same final products and regeneration of the thionine ground state. The difference will be only in the photochemically produced intermediate species.[90] Note that the photochemical oxidation of NADH to $NAD^+$ is an irreversible process.

It should also be noted that direct photoexcitation of NADH (particularly when non-filtered white light is applied) is also possible resulting in its photochemical oxidation without participation of thionine.[91,92] Keeping this possibility in mind, we performed control experiments with the system illuminated in the absence of thionine. Since we did not find any oxidation of NADH in the absence of thionine under the considered experimental conditions, we can conclude that in the present system the photochemical process is primarily mediated by the photoexcited thionine, confirming the process steps shown in Scheme 1.

In order to consider biochemical processes as mimicking binary gates, we set the logic-**0** and **1** levels to convenient or application-motivated values, here initially 0 and $[ATP]_{max} = [Glc]_{max} = 10$ mM for both inputs, whereas the logic levels of the output and, when applicable, intermediate signals are set by the system's functioning itself. Some of the analysis of the gate-function quality is done in terms of the scaled variables,

$$x = [Glc](0)/[Glc]_{max} , \qquad (1)$$

$$y = [ATP](0)/[ATP]_{max} , \qquad (2)$$

$$z = [NADH](t_g)/[NADH]_{max} . \qquad (3)$$

Here the concentrations for the inputs are at time $t = 0$, whereas those for the output (including its maximum, logic-**1** value) are at the measurement gate time, $t_g$. The gate

– 7 –

time, as well as other quantities (including the physical and chemical conditions of the experiment) which are not treated as binary inputs, can be selected for a desirable gate performance. Note that this choice also sets the value of [NADH]$_{max}$ at $t = t_g$.

There have been discussions[48,50,60,93] in the literature involving the analysis of certain measures of the logic-gate quality in terms of this scaled, response-surface function. Networked biomolecular gates are usually noisy (imprecise) and also function in noisy environments (i.e., with fluctuating inputs). Therefore, an important criterion has been avoidance of noise amplification, and preferably having noise suppression which in most cases can be achieved by having the slopes of $z(x,y)$ near all the four logic points less than 1, i.e., $\max_{(x,y) \simeq (0,0),(0,1),(1,0),(1,1)} |\vec{\nabla} z(x,y)| < 1$. The first processing step in our system (Scheme 1) involves a relatively fast enzyme, HK. As will be substantiated in Section 5, for the relevant gate times considered in our data analysis of the final output signal (also presented in Section 5), if we were to define a binary-range variable $s \equiv z_{\text{Glc6P}}$ similarly to Equation (3) but in terms the "intermediate signal" chemical Glc6P, then to a good approximation the response function for the HK-step would be given by a simple expression

$$s \equiv z_{\text{Glc6P}} = \min(x, y) . \tag{4}$$

This function is sketched in Scheme 2, and, interestingly, not only that it realizes the **AND** gate, but it has also been argued[93] not to amplify noise in the inputs (however, it does not suppress noise).

In applications of biochemical logic the binary functions have to be networked/connected in various ways. In our model system the intermediate chemical is processed to convert it to the optically measured output product. Another important quality criterion for biomolecular logic, also related to avoiding noise effects, on a *relative* scale, has been to generate large-intensity signals as compared to the typical fluctuations due to noise. Therefore, the enzymatic cascade should be carried out approximately to saturation, which here implies long time scales for the choice of $t_g$, of order several 100 s. As typical for most enzyme-catalyzed processes, this step results in a convex (in both inputs) overall response, illustrated in Scheme 2 (and revisited in our data analysis in Section 5), which is known to involve noise amplification.[48]

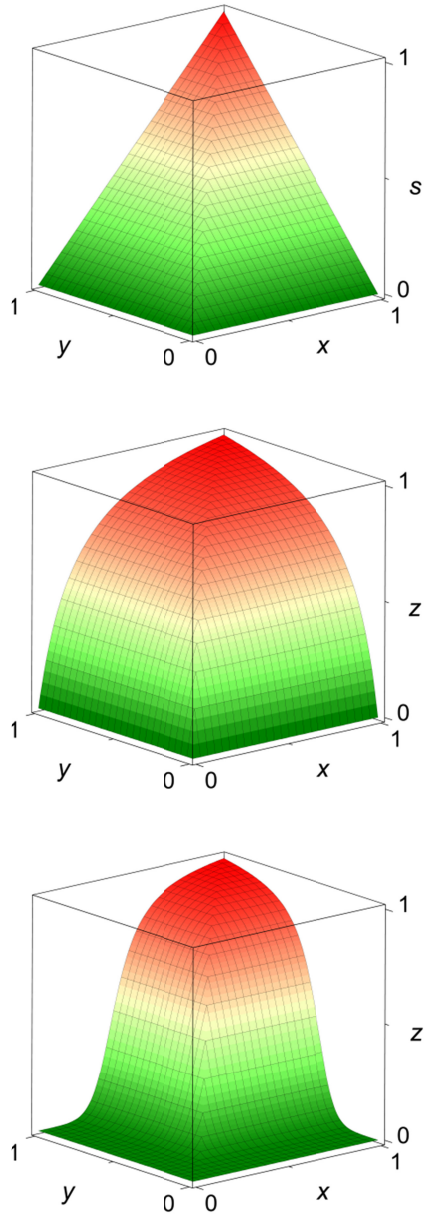

**Scheme 2.** The top panel illustrates the function $s(x,y)$ in Equation (4). The middle and bottom panels illustrate the convex and sigmoid response surfaces $z(x,y) = z(s(x,y))$, which correspond to further processing of the signal as described in the text, with convex $z(s)$, see Equations (11,19), and sigmoid $z(s)$, Equation (17) with Equation (4).

The added filtering,[47,55-65] addressed in the next section, is a process which does not actually change the nature of the signals, but it aims at modifying the system's response from convex to sigmoid, preferably in both of the initial inputs (see Scheme 2), and as a result causing the gate function to suppress noise.[48] However, filtering by the recently developed approaches[47,55-65] can also decrease the overall signal intensity, and therefore careful analysis and design are required for use of such techniques.[60]

**4. Phenomenological Analytical Expression for Biochemical "Signal Filtering"**

In this section we address modeling approaches to describe the dependence $z(s)$, i.e., the signal processing step catalyzed by G6PDH, which in our case is to a good approximation is slow as compared to the earlier, HK, step (as will be discussed in Section 5). As a result of this approximate separation of time scales, the dependence $z(x,y)$ can be introduced via Equation (4), once we have a fitting function to use for $z(s)$. Our aim is to derive a convenient general analytical expression for the process with filtering added, in anticipation that this expression will also be useful in other studies of biochemical filtering. Detailed modeling of all the biocatalytic pathways and other added chemical and physical processes in complicated systems considered for enzymatic information/signal processing is usually impractical. Indeed, the available data are limited, and the process details, especially the enzymatic part (in our case also to a large extent the photochemical part) are not well studied. A complete description would thus involve too many adjustable parameters. Furthermore, such a detailed description is not needed, because we only require a phenomenological parameterization allowing us to explore the shape and slope of the "response surface" function near the four binary "logic points," as well as probe the effects of varying some of the process parameters on it.

An approximate, few-parameter kinetic description is then an option. For example, ignoring added filtering for the moment, let us consider a process catalyzed by an enzyme (here G6PDH) of concentration $E(t)$, where $t$ is time, converting a substrate of concentration $S(t)$, here Glc6P, to a product, $P(t)$, here NADH. We could then write an approximate set of rate equations

$$\frac{dS}{dt} = -k_1 SE, \quad \frac{dP}{dt} = k_2(E_0 - E), \tag{5}$$

$$\frac{dE}{dt} = -k_1 SE + k_2(E_0 - E), \tag{6}$$

corresponding to the simplified kinetics schematically written as

$$S + E \xrightarrow{k_1} C, \quad C \xrightarrow{k_2} E + P, \tag{7}$$

where $C$ is an intermediate compound, and we used $C(t) = E_0 - E(t)$, with $E_0$ defined as the initial value, $E_0 = E(0)$. It is important to again emphasize that this description is already simplified. For example, in our particular case the functioning of G6PDH involves two substrates, and the rates and order of their intake in the formation of complexes may vary.[94-96] Such enzymes have also been noticed to be possibly allosteric.[64,97,98] Details of some of these possibilities and pathways are not known and can depend on the specific enzyme's origin. Furthermore, even in the simplest Michaelis-Menten type picture, we ignored the back reaction in the intermediate compound, $C$, formation. The two-rate-constant approximation has been used in the biochemical information processing studies because of the schematic nature of the response surface modeling and also limited data usually available, as noted earlier. The rate constant parameters $k_{1,2}$ can then be claimed to include other effects indirectly. For example, we expect that $k_1$ varies depending on the initial concentration of the other substrate, $NAD^+$, etc. Such approximate kinetic schemes can also be successfully applied to filtered situations, with another reaction step added.[48,57]

Another extreme of the schematic response-function shape modeling has been to use entirely phenomenological, single- or few-parameter fitting functions.[24,48] An optimal situation would be to also have a certain heuristic understanding how can the fitting parameters be varied to adjust the "gate function" quality and performance, by changing chemical or physical conditions of the system. The Hill equation is the best known example; it, and certain other fitting functions which offer sigmoid shapes have been considered in the literature[66-68] for suitability to fit kinetic data for enzyme functioning. However, these functions are not suitable for the recently studied filtering mechanisms, including the one considered here, which involve "intensity filtering" by redirecting part of the chemical process into another pathway.

To derive a suitable expression, let us first continue with the no-filter kinetics of Equations (5-7) and assume the standard Michaelis-Menten model regime of functioning whereby for most of the process time the reaction in Equation (6) is nearly in steady state (means, the time-derivative is negligibly small). This yields a familiar approximate (even though we use the equal sign) relation

$$E = k_2 E_0 / (k_1 S + k_2) \ . \tag{8}$$

As already mentioned, in biosensing and binary-gates analysis, the back reaction, $C \to S + E$, with rate constant $k_{-1}$, not shown in Equation (7), is ignored partly because these systems are operated in the regime with abundant substrate available, in order to increase the output signal's intensity. Assuming this, we can ignore the depletion of the substrate in an approximate calculation of the derivative of the product, by using the initial substrate value, $S_0 = S(0)$, in the expression for $E$ in Equation (8), in turn used in Equation (5), with the result (another approximate relation):

$$\frac{dP}{dt} = k_1 k_2 E_0 S_0 / (k_1 S_0 + k_2) \ . \tag{9}$$

Either the above product generation rate, or the total product at the gate time, which we write in the form

$$P(t_g) = \frac{k_1 k_2 E_0 t_g}{k_1 + \frac{k_2}{S_0}} \tag{10}$$

for later use, can be considered as the output signal.

Once we divide the output for a general $S_0$ by that at $S_{0,\text{max}}$: $z = (dP/dt) / (dP/dt)_{\text{max}}$, and note that the logic-range variable encountered in Equation (4) is $s = S_0 / S_{0,\text{max}}$, we get a very simple expression

$$z(s) = s(1 + a)/(s + a),  \qquad (11)$$

$$a \equiv k_2/(k_1 S_{0,\text{max}}).  \qquad (12)$$

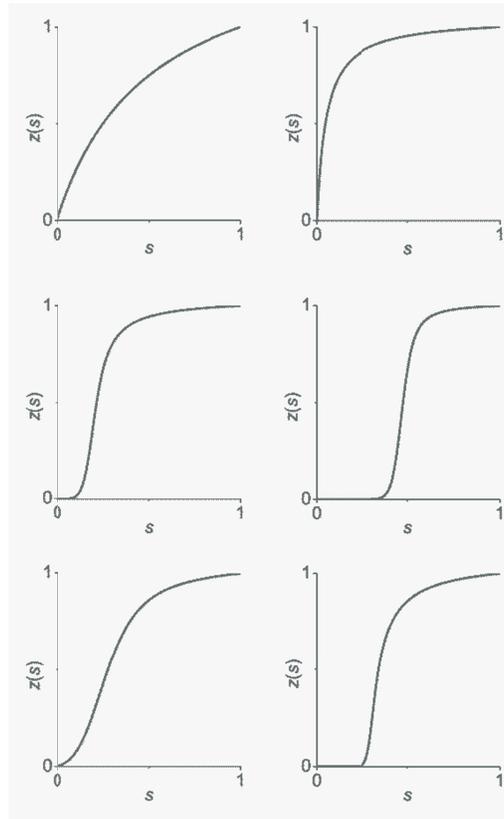

**Scheme 3.** Sketches of the phenomenological fitting functions developed in Section 4 for parameter values randomly selected to offer convenient illustrations. The top panels show the convex function of Equation (11), illustrating, from left to right, the effect of decreasing the value of $a$, i.e., increasing the slope at the origin, cf. Equation (20). The middle panels show the sigmoid function of Equation (17), illustrating, from left to right the effect of increasing the parameter $f$, with the other two parameters, $a$ and $b$, fixed. The bottom panels illustrate the effect of increasing the parameter $b$, with the other two parameters fixed (and not related to the values use in the middle panels).

The expression in Equation (11) was already introduced[24] purely heuristically for studies of non-filtered convex enzymatic signal processing functions, and its two-input generalization was used in the analysis of concatenated "convex-response" **AND** gates.[24] Our present formulation offers a derivation which will shortly be extended to filtered processes. We note that Equation (11) offers a very simple, single-parameter fitting function for convex response, illustrated in Scheme 3 (we will use it in data analysis in Section 5). Its main shortcoming is that it lumps all the dynamical effects into a single constant, and also that its derivation involves many kinetic and mechanistic simplifications and approximations when compared to the full complexity of enzyme-catalyzed processes. Its main advantage is that it offers a simple, closed-form analytical expression. Our present derivation also offers an interpretation of the parameter *a* in terms of kinetic rate constants and chemical concentrations, further explored in Section 5.

Intensity filtering mechanisms recently studied,[47,55-63] have included added chemical reactions that divert some of the input or output into another product which is inactive in signal-processing or represents an intermediate "gate machinery" compound. The approach of the type used here (Scheme 1), involving diversion of some of the output intensity (NADH) by an added process with thionine, converted back into $NAD^+$, can be argued to have two advantages: allowing in some cases to use only a single added process to make the response sigmoid in both inputs (for **AND** gates), and avoiding some of the intensity loss in the total output, by "recycling." We seek a simple phenomenological expression similar to Equation (11), for a filtered (sigmoid) rather than convex process. We note that in writing the approximation Equation (10), we assumed that all the initial input, $S_0$, up to $S_{0,max}$, is available for conversion to the output, and we ignored its depletion for simplicity. However, with filtering, a certain fraction, which we denote $F_0$, of $S_{0,max}$, can potentially be diverted, and the filtering process is usually fast as compared to the main process. In order to include the depletion due to the diversion of some of the input into pathways other than the output signal (directly or via the product: note that filtering is most active at low inputs when the product is proportional to the input) in a schematic way we will approximate the effect of intensity filtering by assuming the simplest kinetics

$$S + F \xrightarrow{k_3} \ldots, \tag{13}$$

where $F$ is initially set to $F_0$. This is obviously a crude approximation aimed at obtaining a very simple fitting form without attention to the details of the actual kinetics. The process (13), if it were ongoing alone, would lead to rate equations

$$\frac{dS}{dt} = \frac{dF}{dt} = -k_3 SF, \tag{14}$$

solved to yield

$$S(t_f) = (S_0 - F_0)S_0 / [S_0 - F_0 e^{-k_3(S_0 - F_0)t_f}], \tag{15}$$

where $t_f$ is the time of the filtering process, which need not be the same as the gate time if the filtering is externally timed, as in our case. We now obtain a schematic expression accounting for the reduced intensity in the output by using the result in Equation (15) instead of $S_0$ on the right-hand side of Equation (10), to get

$$P(t_g) = \frac{k_1 k_2 E_0 t_g}{k_1 + \frac{k_2}{(S_0 - F_0)S_0}[S_0 - F_0 e^{-k_3(S_0 - F_0)t_f}]}. \tag{16}$$

Finally, in terms of the scaled variables we get, after some algebra, cf. Equation (11),

$$z(s) = \frac{s(s-f)\{1 - f + a[1 - fe^{-b(1-f)}]\}}{(1-f)\{s(s-f) + a[s - fe^{-b(s-f)}]\}}, \tag{17}$$

$$f \equiv F_0/S_{0,\max}, \quad b \equiv k_3 S_{0,\max} t_f, \tag{18}$$

which is our proposed phenomenological expression for fitting sigmoid signal shapes obtained with intensity filtering. Parameter $a$ can be estimated from a non-filtered process, whereas the two new parameters introduced, *f* and *b*, quantify the offset of the filtered-response curve, and the sharpness of the transition from the reduced response at low inputs to one returning to convex for larger inputs. These parameters are further discussed in the next section, whereas their effect on the shape of the curve *z(s)* is illustrated in Scheme 3. Generally, we expect that for physically meaningful results the values of the three phenomenological parameters introduced in this section should satisfy $a > 0;\ 0 \leq f < 1;\ b \geq 0$.

## 5. Data Analysis: Results and Discussion

Let us first consider the system without filtering. As mentioned in Section 2, the experiments before filtering were performed up to the gate time of $t_g = 600$ s. Figure 1 shows the measured output for this time as a function of the two inputs. The fitted surface was obtained as a single-parameter, *a*, fit according to Equations (4) and (11) which combine to yield

$$z(x, y) = \min(x, y)(1 + a)/[\min(x, y) + a] . \tag{19}$$

Of course, the actual data fitting was done in terms of the physical quantities rather than the scaled variables, cf. Equations (1-3). Figure 1 also shows the fitting of the data if we select shorter gate times, $t_g = 150$ or $50$ s. In fact, similar fitting was carried out for several other times, and the results for the parameter *a* for several $t_g$ values are reported in Figure 2, where for later use we actually plotted the values of the slope of the function $z(s)$ in Equation (11) at $s = 0$, which is related to *a*:

$$\left(\frac{dz}{ds}\right)_{s=0} = 1 + \frac{1}{a} . \tag{20}$$

We note that the proposed fitting functions for the non-filtered system work relatively well: the data are approximately symmetrical in terms of the two inputs, and the quality of the fits is rather good considering that the data are noisy and that this is actually only a single-parameter fitting. Before we address the implications of this observation for the kinetics of the constituent processes, let us discuss the noise handling of this non-filtered **AND** gate realization.

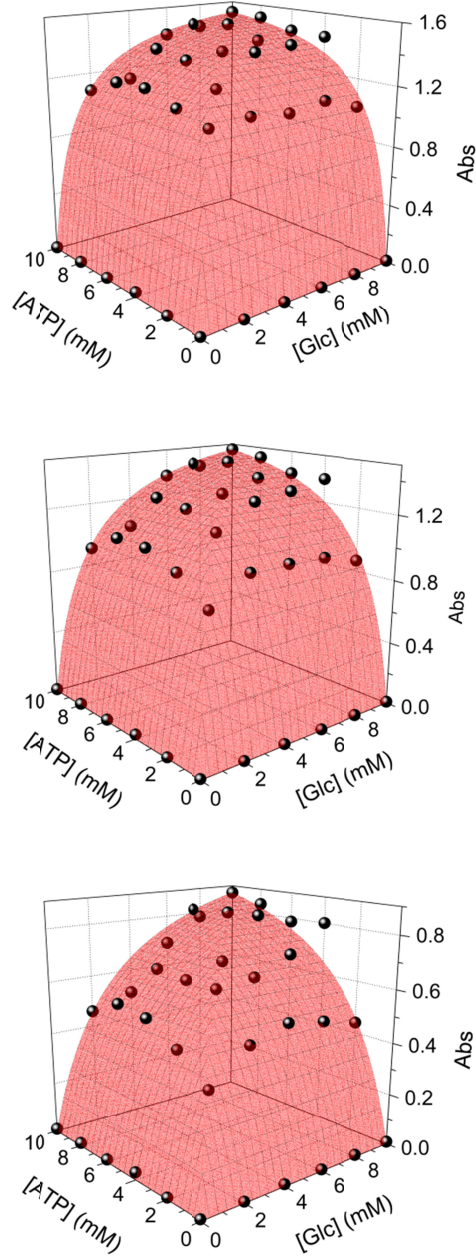

**Figure 1.** Top to bottom: data for the absorbance (Abs), plotted as heavy points, for gate time selections of $t_g$ = 600, 150 and 50 s, respectively, for the process without filtering. The surfaces represent the single-parameter fits according to Equation (19). (The point symbols above the surface are in black; those below it are in dark carmine color.)

Inspection of Figure 1 illustrates the main reason for preferring larger gate times (600 s in this case). Indeed, while the noisiness of the gate realization is comparable for the shown gate times, the overall signal intensity is larger for longer times, markedly so if we compare for instance the 50 s and 600 s data shown. Thus, the *relative* fluctuations in the output due to this source of noise are smaller at larger gate times. Noise is typically generated not only in the functioning of enzymatic processes, but it is also present in the input signals. In applications such as biomedical sensing/diagnostics, the inputs are not accurately defined as "binary" values. Rather, they are typically distributed, with the distributions differing for instance, for normal physiological vs. pathophysiological conditions.[40] At any stage of the information/signal processing, noise can be added due to the processing steps, but more importantly, it is also transmitted from the input to output in each step. For large-scale networking, such "analog" noise amplification should be avoided or minimized.[48,50] Preferably, most signal processing steps should actually suppress analog noise at all the logic points.[48] As mentioned earlier, in many cases the slopes of the response surface in the vicinity of the logic points provide a good measure of the analog noise transmission.

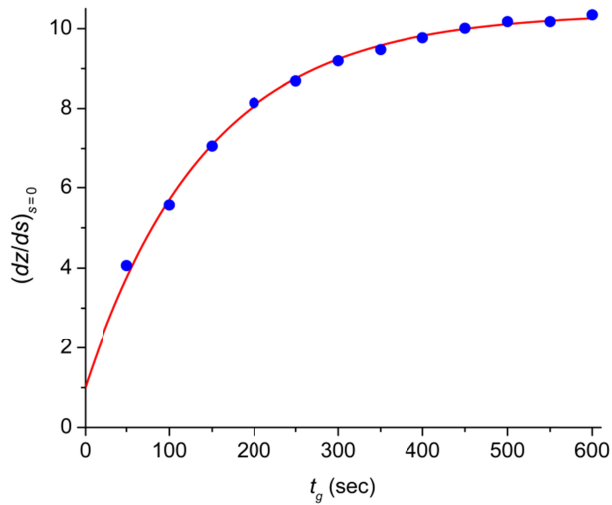

**Figure 2.** The symbols represent the values of the slope in Equation (20), calculated from the fitted *a* values for gate times of 50, 100, …, 600 s. The curve was fitted according to Equation (21).

The fact that a fitting-function surface has a ridge can be shown[93] to be immaterial in such considerations. Actually, in our case the function in Equation (4) has near-logic-point slope values of 1, means no noise amplification or suppression. Therefore, slopes of the function $z(s)$ alone, near the logic points **0** and **1** determine noise transmission of the whole **AND** gate. The slope at $s = 1$ happens to be the inverse of the value in Equation (20), and is therefore always smaller than 1. However, as shown in Figure 2 and Scheme 3, for the non-filtered case the slope at $s = 0$ exceeds 1. Thus, overall, the **AND** gate without filtering amplifies noise on its transmission from the inputs (at the small input values) to the output, and the amplification factor increases from somewhat over 100% (somewhat over 1) for shorter gate times, to order 1000% (tenfold amplification) at the desirable gate time of 600 s. These considerations clearly illustrate the need for the added filtering process.

Regarding the applicability of the convex, Equation (11), and other phenomenological fitting expressions, we also need to address the issue of how literally should the expressions for the coefficients, such as Equation (12), be taken? The fitting functions are phenomenological and can only be used for describing the response shape. They should not be treated in any way as accurately or even semi-accurately representing the kinetics of the enzymatic processes. However, relations such as Equation (12) can provide additional useful information, as well as suggest how to optimize the gate functioning. The assumed fast nature of the process step biocatalyzed by HK, which is obviously an approximation, can be studied in this context. In such regime, the functioning of this enzyme will not be steady state, but rather can be assumed to be kinetically limited by the availability (the remaining unused concentration) of the smaller of the two inputs, Glc or ATP. Let us denote the time-dependent concentration of this input as $I(t)$, and assume the simplest first-order process, $dI/dt = -k_I I(t)$. Then the amount of the consumed input is $I(0)(1 - e^{-k_I t})$, which then also describes the gate-time dependence of the parameter $S_{0,\text{max}}$ in Equation (12). Up to an unknown constant, this offers a parameterization for the gate-time dependence of $a$ as proportional to $1/(1 - e^{-k_I t_g})$, which we rewrite for the slope, see Equation (2) and Figure 2,

$$\left(\frac{dz}{ds}\right)_{s=0}(t_g) = 1 + \left[\left(\frac{dz}{ds}\right)_{s=0}(\infty) - 1\right](1 - e^{-k_I t_g}) . \tag{21}$$

As already noted in Figure 2, the slope increases from 1 for short process (gate) times to the large-time ($t_g = \infty$) value which is much larger than 1; the solid line in the figure was fitted to yield the estimates

$$\left(\frac{dz}{ds}\right)_{s=0}(\infty) \simeq 10.4 , \quad k_I \simeq 6.97 \times 10^{-3} \text{ s}^{-1} . \tag{22}$$

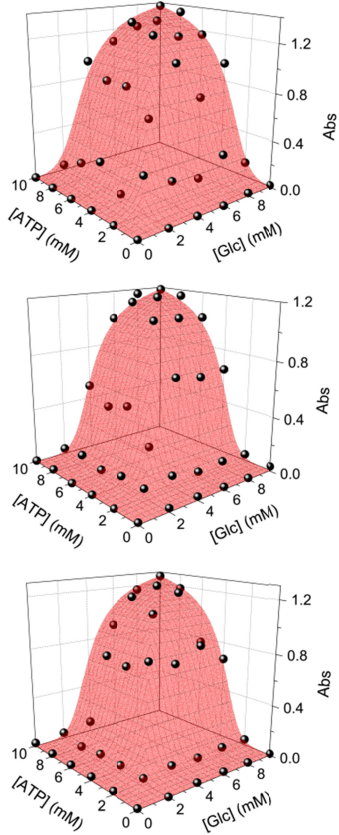

**Figure 3.** Top to bottom: data at $t_g = 900$ s, for the experiments with filtering activated during 600 through 900 s, with light intensity and thionine concentration of 9000 lux and 200 μM; 9000 lux and 250 μM; and 12000 lux and 200 μM, respectively, plotted as heavy points. The surfaces represent the two-parameter fits according to Equation (17), with Equation (4), and with the third parameter, $a$, set to its large-time value (obtained earlier from the non-filtered data), cf. Equations (20,22). The fitted parameter values were, top to bottom, $f$, $b$ = 0.29, 18.9; 0.37, 22.0; and 0.33, 25.5. (The point symbols above the surface are in black; those below it are in dark carmine color.)

We note that the time scale associated with the rate-constant estimate, $1/k_I \sim 140$ s, is small but not dramatically so, as compared to the preferred gate times, 600 s (and 900 s with filtering). Our assumption of the separation of time scales of the two enzymatic steps is thus valid, but only approximately.

Figure 3 illustrates the results for the filtered system. The filtering process was active for times from 600 s to the final gate time of 900 s. The data obtained for light intensity and thionine concentration of 9000 lux and 200 μM, shown in the top panel of the figure, exemplify the achieved effect: The response surface is now sigmoid with respect to both inputs, with small slopes (means, noise suppression on its transmission from the inputs to output, quantified shortly) at all the four logic points of the **AND** gate function. The price paid is, as expected, the loss of the overall signal intensity (compare to the logic-point **1,1** value of the signal in the top panel in Figure 1). The other two panels in Figure 3 illustrate that increasing the "activity" of the filtering process by either using more thionine (middle panel) or increased light intensity (bottom panel) can push the inflection region of the sigmoid shape to larger inputs: increase the fitted parameter $f$ (recall Scheme 3), and also make the bottom part of the inflection region somewhat sharper: increase $b$ (cf. Scheme 3), but at the expense of having an additional loss of the physical signal intensity. These variations in the sigmoid shape are expected, because of the increase of the parameters $F_0$ and $k_3$ entering the expressions in Equation (18). However, we caution that such considerations are at best qualitative, because of the phenomenological nature of the developed sigmoid fitting shape, Equation (17).

The noise-transmission properties of the filtered **AND** gate for our preferred set of parameters, are quantified in Figure 4. The regions of the slope less than 1, corresponding to noise suppression, are found near all the four logic points, and extend at least approximately 15% of the overall signal span. The loss of the overall signal intensity due to filtering in this case is of a comparable order of magnitude, about 10%, and is acceptable because the improved noise tolerance achieved.

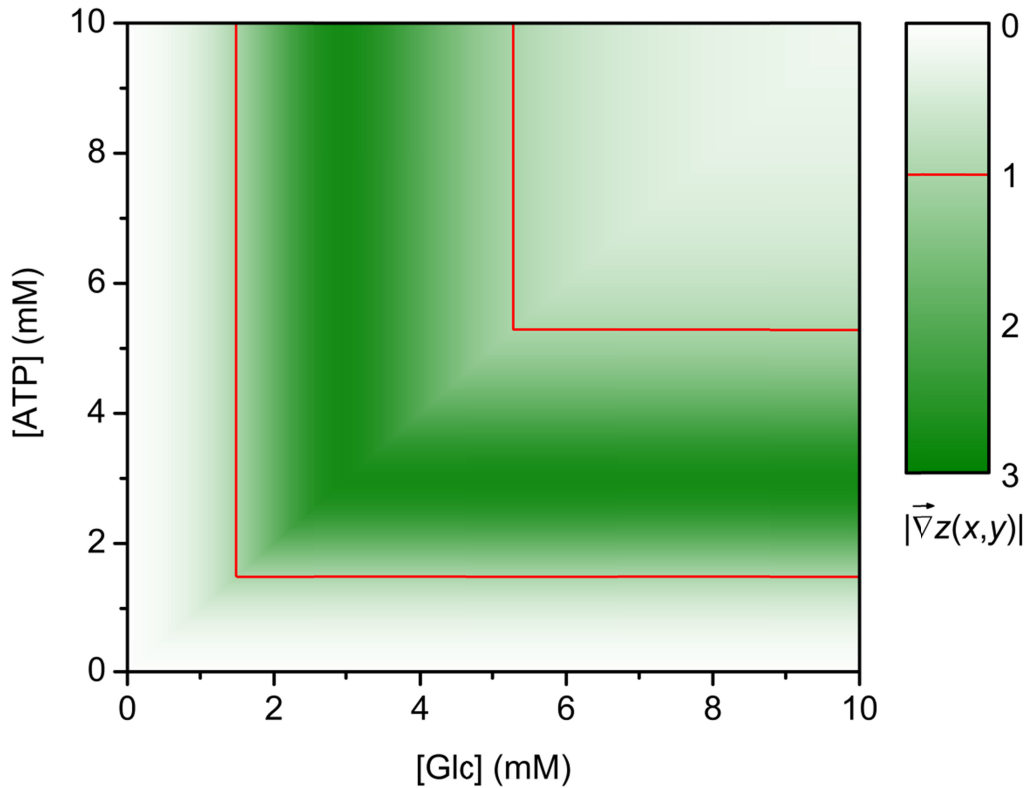

**Figure 4.** The slope of the fitted sigmoid gate-response function in terms of the scaled logic-range variables, $|\vec{\nabla}z(x,y)|$, plotted for the filtered system with light intensity of 9000 lux and thionine concentration of 200 µM. The red lines mark the slope values 1, delineating the regions of the inputs-to-output noise suppression, which are the lighter-shaded regions at the "pedestal" and "roof" of the sigmoid shape shown in the top panel of Figure 3.

## 6. Conclusions

This work reported a novel approach to accomplishing a two-input gate with sigmoid response in both inputs. A relatively fast processing of the two inputs to yield an intermediate signal, was accompanied by further processing of this signal which, when supplemented with filtering, resulted in the desired sigmoid overall gate function response. While not as well-defined as "clocking" in electronics, the process here involved a sequence of timed steps, and also utilized an external physical (optical) control. The studied system was specially designed for the recycling NADH to $NAD^+$.

For other cofactors/products different systems should be considered. It should be noted however, that the NADH/NAD$^+$ cofactors are used by many different enzymes, thus making the studied system applicable in many different enzyme-based logic systems.

Our new theoretical expression, Equation (17), for fitting sigmoid functions in situations involving "intensity filtering," tested well with the present data and should be useful for modeling other systems, as well as, when supplemented with the expressions for the parameters, Equations (12) and (18), can provide a phenomenological approach to adjusting the chemical and physical conditions to improve the system performance.

**Acknowledgements**

We acknowledge support of this research by the National Science Foundation under grants CBET-1066397 and CCF-1015983.

– From Biomolecular Logic Gates to Animal Model. *Analyst* **2012**, *137*, 1768–1770.

(39) Zhou, N.; Windmiller, J. R.; Valdés Ramírez, G.; Zhou, M.; Halámek, J.; Katz, E.; Wang, J. Enzyme-Based NAND Gate for Rapid Electrochemical Screening of Traumatic Brain Injury in Serum. *Anal. Chim. Acta* **2011**, *703*, 94–100.

(40) Zhou, J.; Halámek, J.; Bocharova, V.; Wang, J.; Katz, E. Bio-Logic Analysis of Injury Biomarker Patterns in Human Serum Samples. *Talanta* **2011**, *83*, 955–959.

(41) Halámek, J.; Bocharova, V.; Chinnapareddy, S.; Windmiller, J. R.; Strack, G.; Chuang, M.-C.; Zhou, J.; Santhosh, P.; Ramirez, G. V. Arugula, M. A.; et al. Multi-Enzyme Logic Network Architectures for Assessing Injuries: Digital Processing of Biomarkers. *Mol. Biosyst.* **2010**, *6*, 2554–2560.

(42) Melnikov, D.; Strack, G.; Zhou, J.; Windmiller, J. R.; Halámek, J.; Bocharova, V.; Chuang, M.-C.; Santhosh, P.; Privman, V.; Wang, J.; et al. Enzymatic AND Logic Gates Operated under Conditions Characteristic of Biomedical Applications. *J. Phys. Chem. B* **2010**, *114*, 12166–12174.

(43) Halámek, J.; Windmiller, J. R.; Zhou, J.; Chuang, M.-C.; Santhosh, P.; Strack, G.; Arugula, M. A.; Chinnapareddy, S.; Bocharova, V.; Wang, J.; et al. Multiplexing of Injury Codes for the Parallel Operation of Enzyme Logic Gates. *Analyst* **2010**, *135*, 2249–2259.

(44) Chuang, M.-C.; Windmiller, J. R.; Santhosh, P.; Valdés-Ramírez, G.; Katz, E.; Wang, J. High-Fidelity Simultaneous Determination of Explosives and Nerve Agent Threats via Booleam Biocatalytic Cascade. *Chem. Commun.* **2011**, *47*, 3087–3089.

(45) Bocharova, V.; Halámek, J.; Zhou, J.; Strack, G.; Wang, J.; Katz, E. Alert-Type Biological Dosimeter Based on Enzyme Logic System. *Talanta* **2011**, *85*, 800–803.

(46) Stojanovic, M. N.; Stefanovic, D. Chemistry at a Higher Level of Abstraction. *J. Comput. Theor. Nanosci.* **2011**, *8*, 434–440.

(47) Domanskyi, S.; Privman, V. Design of Digital Response in Enzyme-Based Bioanalytical Systems for Information Processing Applications. *J. Phys. Chem. B* **2012**, *116*, 13690–13695.
– 27 –

(58) Rafael, S. P.; Vallée-Bélisle, A.; Fabregas, E.; Plaxco, K.; Palleschi, G.; Ricci, F. Employing the Metabolic "Branch Point Effect" to Generate an All-or-None, Digital-Like Response in Enzymatic Outputs and Enzyme-Based Sensors. *Anal. Chem.* **2012**, *84*, 1076–1082.

(59) Pita, M.; Privman, V.; Arugula, M. A.; Melnikov, D.; Bocharova, V.; Katz, E. Towards Biochemical Filter with Sigmoidal Response to pH Changes: Buffered Biocatalytic Signal Transduction. *Phys. Chem. Chem. Phys.* **2011**, *13*, 4507–4513.

(60) Privman, V.; Halámek, J.; Arugula, M. A.; Melnikov, D.; Bocharova, V.; Katz, E. Biochemical Filter with Sigmoidal Response: Increasing the Complexity of Biomolecular Logic. *J. Phys. Chem. B* **2010**, *114*, 14103–14109.

(61) Vallée-Bélisle, A.; Ricci, F.; Plaxco, K. W. Engineering Biosensors with Extended, Narrowed, or Arbitrarily Edited Dynamic Range. *J. Am. Chem. Soc.* **2012**, *134*, 2876−2879.

(62) Kang, D.; Vallée-Bélisle, A.; Plaxco, K. W.; Ricci, F. Re-engineering Electrochemical Biosensors To Narrow or Extend Their Useful Dynamic Range. *Angew. Chem. Int. Ed.* **2012**, *51*, 6717–6721.

(63) Ricci, F.; Vallée-Bélisle, A.; Plaxco, K. W. High-Precision, In Vitro Validation of the Sequestration Mechanism for Generating Ultrasensitive Dose-Response Curves in Regulatory Networks. *PLoS Comput. Biol.* **2011**, *7*, article #e1002171

(64) Privman, V.; Pedrosa, V.; Melnikov, D.; Pita, M.; Simonian, A.; Katz, E. Enzymatic AND-Gate Based on Electrode-Immobilized Glucose-6-phosphate Dehydrogenase: Towards Digital Biosensors and Biochemical Logic Systems with Low Noise. *Biosens. Bioelectron.* **2009**, *25*, 695–701.

(65) Pedrosa, V.; Melnikov, D.; Pita, M.; Halámek, J.; Privman, V.; Simonian, A.; Katz, E. Enzymatic Logic Gates with Noise-Reducing Sigmoid Response. *Int. J. Unconv. Comput.* **2010**, *6*, 451–460.

(66) Kurganov, B. I.; Lobanov, A. V.; Borisov, I. A.; Reshetilov, A. N. Criterion for Hill Equation Validity for Description of Biosensor Calibration Curves. *Anal. Chim. Acta* **2001**, *427*, 11–19.

– 29 –